\documentclass[aps,prd,twocolumn,superscriptaddress,showpacs,bibnotes,preprintnumbers,amsmath,amssymb]{revtex4}
\usepackage{graphicx}
\usepackage{dcolumn}
\usepackage{bm}
\usepackage{epstopdf}
\usepackage{mathtools}
\usepackage{natbib}
\usepackage{captcont}

\def\apj{Astrophys. J.}
\def\prd{Phys. Rev. D}
\def\mnras{Mon. Not. R. Astron. Soc.}
\def\jcap{J. Cos. Astropart. Phys.}
\def\aap{Astron. Astrophys.}
\def\araa{Ann. Rev. Astron. Astrophys.}
\def\aj{Astronomical Journal}

\begin{document}
\title{Diffuse PeV neutrino emission from Ultra-Luminous Infrared Galaxies}

\affiliation{Key Laboratory of Dark Matter and Space Astronomy, Purple Mountain Observatory, Chinese Academy of Sciences, Nanjing 210008, China }
 \affiliation{School of Astronomy and Space Science,, Nanjing University, Nanjing, 210093, China}
 \affiliation{Key laboratory of Modern Astronomy and Astrophysics(Nanjing University), Ministry of Education, Nanjing 210093, China}

\author{Hao-Ning He}
\affiliation{Key Laboratory of Dark Matter and Space Astronomy, Purple Mountain Observatory, Chinese Academy of Sciences, Nanjing 210008, China }
 \author{Tao Wang}
 \affiliation{School of Astronomy and Space Science,, Nanjing University, Nanjing, 210093, China}
 \affiliation{Key laboratory of Modern Astronomy and Astrophysics(Nanjing University), Ministry of Education, Nanjing 210093, China}
 \author{Yi-Zhong Fan$^\ast$}
\affiliation{Key Laboratory of Dark Matter and Space Astronomy, Purple Mountain Observatory, Chinese Academy of Sciences, Nanjing 210008, China }
 \author{ Si-Ming Liu}
\affiliation{Key Laboratory of Dark Matter and Space Astronomy, Purple Mountain Observatory, Chinese Academy of Sciences, Nanjing 210008, China }
 \author{Da-Ming Wei}
\affiliation{Key Laboratory of Dark Matter and Space Astronomy, Purple Mountain Observatory, Chinese Academy of Sciences, Nanjing 210008, China }

\begin{abstract}
Ultra-luminous infrared galaxies (ULIRGs) are
the most luminous and intense starburst galaxies in the Universe.
Both their star-formation rate (SFR) and gas surface mass density
are very high, implying a high supernovae rate and an
efficient energy conversion of energetic protons. A small fraction of these supernovae is the so-called hypernovae with a typical kinetic energy
$\sim 10^{52}$ erg and a shock velocity $\geq 10^{9}~{\rm cm~s^{-1}}$. The strong shocks driven by hypernovae are able to accelerate cosmic ray protons up to $10^{17}$ eV. These energetic protons lose a good fraction of their energy
through proton-proton collision when ejected into
very dense interstellar medium, and as a result,
produce high energy neutrinos ($\leq 5$ PeV).
Recent deep infrared surveys provide solid constraints on
the number density of ULIRGs across a wide redshift range
$0 \leqslant z \leqslant 2.3$, allowing us to derive the flux of diffuse neutrinos
from hypernovae.
We find that at PeV energies, the diffuse neutrinos contributed by ULIRGs
are comparable with the atmosphere neutrinos with the flux of $2\times10^{-9}{\rm GeV cm^{-2}s^{-1} sr^{-1}}$,
by assuming the injected cosmic ray spectrum as
$dN'_{\rm p}/d\varepsilon'_{\rm p}\propto \varepsilon_{\rm p}^{\prime-2}$.
\end{abstract}
\maketitle

\section{Introduction}
The galactic supernova remnants (SNRs) are widely suggested to be
the dominant source for the cosmic rays (CRs) at energies below the ``knee"
at $\sim 3\times 10^{15}\rm eV$, most probably through the diffusive shock
acceleration mechanism \cite{Hillas2005}.
Though the details are still to be figured out, it is generally believed that the maximum energy of CRs accelerated by SNRs depends on
both the velocity and the kinetic energy of the supernova outflow. A small fraction of the supernovae has a typical kinetic energy
$E_{\rm k}\sim 0.5-5\times 10^{52}$ erg and a typical velocity $V \sim 10^{9}~{\rm cm~s^{-1}}(E_{\rm k}/10^{52}~{\rm erg})^{1/2}(M_{\rm SN}/10M_\odot)^{-1/2}$, both are substantially larger than that of normal supernova, where $M_{\rm SN}$ is the rest mass of the SN ejecta.
These peculiar supernovae, such as SN 1997ef, SN 1997dq, SN 1998bw and SN 2002ap, have been called the hypernovae \cite{Iwamoto1998,Mazzali2002, Mazzali2004}.
The maximum energy of the protons accelerated at the shock front of a supernova expanding into the uniform dense interstellar
medium (ISM) can be estimated as
$\varepsilon'_{\rm p,max}\approx 10^{17}{\rm eV}({V\over 10^{9}~{\rm cm~s^{-1}}})^2
	({n\over 10^3\rm cm^{-3}})^{1/6}
	({M_{\rm SN}\over 10M_\odot})^{1/3}$,
where $n$ is the number density of ISM \cite{Bell2001}.
Such a fact motivates some colleagues to suggest that hypernova remnants are the dominant source of cosmic rays above the knee and the cosmic ray spectrum/flux up to $\sim10^{18}-10^{19}$ eV may be accounted for as long as the variety of supernovae has been taken into account \cite{Dermer2001,Sveshnikova2003,Wang2007,Budnik2008,Fan2008}.

Ultra-luminous infrared galaxies (ULIRGs), first discovered in large numbers by the Infrared Astronomical Satellite in 1983, are among the most luminous objects in the
local universe with infrared luminosity $L_{\rm 8-1000 \mu m}>10^{12}L_{\odot}$ \cite{Sanders1988,Sanders1996}.
The large infrared luminosities are attributed to large amounts of dust, which absorb ultraviolet (UV) photons and re-radiate them in the infrared (IR). Comprehensive observations show that ULIRGs are powered mainly by a large population of hot young stars, i.e., a ``starburst", with a significant fraction also containing an IR-luminous AGN \cite{Lonsdale2006}. Local ULIRGs are exclusively mergers of gas rich galaxies accompanied by concentrated dust-enshrouded starburst, with very high star formation rate (SFR) $\gtrsim 200 {\rm M_{\odot}~yr^{-1}}$. The supernova rate is expected to be high and huge amounts of cosmic ray particles are accelerated.  Moreover, the relatively small sizes $\sim 1~{\rm kpc}$ of ULIRGs mean that the ISM density is also quite high, i.e., $<n>\simeq 10^3-10^4{\rm cm^{-3}}$\cite{Downes1998},
suggesting that the accelerated high energy protons produced in ULIRGs have a great chance to interact with the interstellar medium nucleons, to produce pions, and decay into secondary electrons and positrons, $\gamma$-rays, and neutrinos.
If the energy loss time of the protons through the proton-proton collisions is shorter than the starburst lifetime and
the confinement time, the protons will lose most of their energy before escaping and produce interesting observational signals \cite{Pohl1994,Lacki2011}.

Previously the possible high energy neutrino emission from the starburst galaxies has been investigated by assuming that the observed GeV photons are produced via the decay of pions \cite{LoebWaxman2006,Stecker2007}. The events are expected to be significantly enhanced if the nearby starburst galaxies host some Gamma-ray Bursts (GRBs) \cite{Becker2009}.
In this work, we focus on the PeV neutrino emission produced via the interaction of the accelerated Cosmic Rays with the dense environment in the ULIRGs.

\section{The Hypernovae Rate in the ULIRGs}

Though ULIRGs are very rare in the local universe, they are vastly more numerous at high redshifts. The relative contribution of ULIRGs to the SFR density of the universe also increases with redshift, and may even be the dominant component at $z \sim 2$ \cite{Dale2002}. Thus it is essential to study the role of ULIRGs across a wide redshift range in producing PeV neutrinos.
Observational data shows that core collapse supernovae of type SN-Ib/c contribute
with $11\%$ of the total SN rate \cite{Cappellaro1999}.
The hypernovae rate to the normal Ib/c SN rate is $7\%$ in the local universe \cite{Guetta2007}.
Therefore the ratio of the hypernovae rate
to the supernovae rate can be estimated to be $f_{\rm HS}\simeq0.01$.

Recent IR observations show that the SFR density for ULIRGs
increases rapidly at $z<1$,
and stays as a constant at higher redshift $1.2\leqslant z\leqslant2.3$,
which can be approximated as \cite{Magnelli2011}
\begin{equation}
\rho_{\rm SFR}(z)=\left\{{}
\begin{array}{ll}
1.9\times10^{-4}e^{3.3z}{\rm M_\odot yr^{-1} Mpc^{-3}},\,\,\,\,z<1.2;\\
0.01{\rm M_\odot yr^{-1} Mpc^{-3}},\,\,\,\,\,\,\,\,\,\,\,\,\,\,\,\,\,\,1.2\leqslant z\leqslant2.3.
\end{array}\right.
\end{equation}

The supernova rate is related to the SFR via \cite{Fukugita2003}
\begin{equation}
R_{\rm SN}(z)=1.2\times10^{-2}\rho_{\rm SFR}(z)/M_{\odot}.
\end{equation}
Therefore, the hypernovae rate is
\begin{eqnarray}
R_{\rm HN}(z)&=&f_{\rm HS}R_{\rm SN}(z)\\\nonumber
&=&\left\{{}
\begin{array}{ll}
2.4\times 10^{-8}e^{3.3z}{\rm yr^{-1} Mpc^{-3}}, \,\,\,\,z<1.2;\\
1.2\times 10^{-6}{\rm yr^{-1} Mpc^{-3}},\,\,\,\,1.2\leqslant z\leqslant2.3.
\end{array}\right.
\end{eqnarray}
Taking into account cosmology, the hypernovae number occurring
in ULIRGs with redshift $\leqslant z$ per year is numerically calculated by
\begin{equation}
N_{\rm HN}(z)=\frac{c}{H_0}\int_0^zdz\frac{R_{\rm HN}(z)4\pi D_{\rm c}(z)^2}
{\sqrt{\Omega_{\rm \Lambda}+(1+z)^3\Omega_{\rm m}}},
\end{equation}
where $D_{c}(z)=\frac{c}{H_0}\int_0^zdz\frac{1}{\sqrt{\Omega_{\rm \Lambda}+(1+z)^3\Omega_{\rm m}}}$ is the comoving distance, $c$ is the light speed.
Hereafter, we adopt the Hubble constant as $H_0=70{\rm km s^{-1} Mpc^{-1}}$,
the matter density as $\Omega_{\rm m}=0.3$, and the dark energy density as $\Omega_{\Lambda}=0.7$ in the flat universe.

\section{Diffuse PeV neutrino emission from ULIRGs.}
The protons accelerated by the hypernovae
would lose energy into the $\gamma$-ray photons, electrons and positrons,
and neutrinos, through proton-proton collisions when injected into
the interstellar mediums.
A part of the protons energy will convert into neutrinos via the decay of charged pions,
$\pi^+\rightarrow \mu^++\nu_\mu\rightarrow e^++\nu_e+\bar\nu_\mu+\nu_\mu$
and $\pi^-\rightarrow \mu^-+\bar\nu_\mu\rightarrow e^-+\bar\nu_e+\bar\nu_\mu+\nu_\mu$.

The energy loss time of protons is
$\tau_{\rm loss}=(0.5n\sigma_{\rm pp}c)^{-1}$,
where the factor $0.5$ is inelasticity \cite{Gaisser1990},
and $\sigma_{\rm pp}$ is the inelastic nuclear collision cross section,
which is $\gtrsim 70 {\rm mb}$ for the protons at energies $\varepsilon'_{\rm p}\geq 10$ PeV that is of our great interest.
Introducing a parameter $\Sigma_{\rm gas}\equiv m_{\rm p}nl$
as the surface mass density of the gas,
with $l$ as the scale of the dense region in the galaxies,
the energy loss time reads \cite{Condon1991}
\begin{equation}
\tau_{\rm loss}=1.4\times 10^{4}{\rm yr}\frac{l}{100\rm pc}\left(\frac{\Sigma_{\rm gas}}{1.0~{\rm g ~cm^{-2}}}\right)^{-1}.
\end{equation}

Gas surface density can be derived from their SFR density based on an empirical relation that SFR surface density scales as some positive power $\beta \sim 1.4$ of the
local gas surface density, i.e., the Kennicutt-Schmidt law \cite{Kennicutt1998}.
\begin{equation}
\Sigma_{\rm SFR}=(2.5\pm 0.7)\times10^{-4}
\left(\frac{\Sigma_{\rm gas}}{1 \rm M_\odot pc^{-2}}\right)^{1.4\pm0.15}
{\rm M_\odot yr^{-1} kpc^{-2}}.
\end{equation}
ULIRGs have SFR $\gtrsim 200 M_{\odot} {\rm yr^{-1}}$\cite{Soifer2000}.
If we assume a half-light radius of $\sim 1 ~{\rm kpc}$,
it yields a SFR density $\gtrsim 32~M_{\odot}~{\rm yr^{-1}~ kpc^{-2}}$,
consistent with observations of ULIRGs at both low and high redshifts.
According to the Kennicutt-Schmidt law, the derived gas surface density is
$\gtrsim 1.0~{\rm g~cm^{-2}}$. Hence,
the energy loss time for the known ULIRGs is much shorter than the starburst lifetime in ULIRGs, which ranges from $10^7$ to $10^8$ years \cite{Solomon2005}.

Another important factor to determine the fraction of protons energy conversion
is the magnetic confinement time of protons.
Adopting the similar confinement time as the Galaxy,
and simply considering that the confinement time depends on
the proton's Larmor radius($\propto \varepsilon'_{\rm p}/B$),
the confinement time of the proton with energy $\varepsilon'_{\rm p}$ can be estimated as \cite{LoebWaxman2006}
\begin{equation}
\tau_{\rm conf}\approx 10^4{\rm yr}\left(\frac{\varepsilon'_{\rm p}}{10\rm PeV}\right)^{-0.5}\left(\frac{B}{\rm 3\mu G}\right)^{0.5}.
\end{equation}

It is found in Thompson et al. (2006) \cite{Thompson2006} that the magnetic field strength of the starburst galaxies
scales linearly with $\Sigma_{\rm gas}$, i.e.,
$B\approx 3{\rm \mu G}\frac{\Sigma_{\rm gas}}{2.5\times10^{-3}\rm gcm^{-2}}$. Consequently,
the confinement time can be rewritten as
\begin{equation}
\tau_{\rm conf}\approx 2\times10^5{\rm yr}\left(\frac{\varepsilon'_{\rm p}}{10\rm PeV}\right)^{-0.5}(\frac{\Sigma_{\rm gas}}{1.0\rm g cm^{-2}})^{0.5}.
\end{equation}
The fraction of the energy that the protons lose into pions is $f_\pi=1-\exp(-\tau_{\rm conf}/\tau_{\rm loss})$, which is close to 1 as long as
$\tau_{\rm conf}\geqslant\tau_{\rm loss}$.
As a result, the protons with energy $\varepsilon'_{\rm p}$
lose almost all of their energy via the proton-proton collision
before escaping from the starburst galaxies
as long as $\tau_{\rm loss}\leq \tau_{\rm conf}$,
which constrains the critical gas surface density $\Sigma_{\rm crit}$ as
\begin{equation}
\Sigma_{\rm gas}\gtrsim\Sigma_{\rm crit}=0.17~{\rm g~cm^{-2}}
\left(\frac{\varepsilon'_{\rm p}}{10\rm PeV}\right)^{1/3}\left(\frac{l}{100\rm pc}\right)^{2/3}.
\label{eq:Sigma_gas}
\end{equation}
Consequently, for the ULIRGs with the gas surface density
$\Sigma_{\rm gas}\gtrsim 1.0~{\rm g~cm^{-2}}$,
the CR protons with energy up to $\sim
2.0\times10^3{\rm PeV}\left(\frac{l}{100{\rm pc}}\right)^{-2}$
will lose almost all their energy via interacting with the dense ISM.

The charged pion, whose energy is $\varepsilon'_\pi=0.2\varepsilon'_p$,
will then decay to produce four leptons, which share the energy equally.
Therefore, the fraction of the protons energy converted into neutrinos is
$\epsilon_{\nu}=0.05$ \cite{Kelner2006}.
The observed neutrinos in the energy range
($\varepsilon_{\nu,1}, \varepsilon_{\nu,2}$)
are produced by the protons in the energy range
($\varepsilon'_{p,1}, \varepsilon'_{p,2}$) in the ULIRGs,
where $\varepsilon'_{p,1}=(1+z)\varepsilon_{\nu,1}/\epsilon_\nu$
and $\varepsilon'_{p,2}=(1+z)\varepsilon_{\nu,2}/\epsilon_\nu$.
The energy fraction of the protons producing the neutrinos with energy
between $\varepsilon_{\nu,1}$ and $\varepsilon_{\nu,2}$ is (for $\alpha>2.0$) 
\begin{equation}
\epsilon'_{\rm dec}
=\frac{\varepsilon_{p,1}^{\prime2-\alpha}-\varepsilon_{p,2}^{\prime2-\alpha}}
{\varepsilon_{p,\rm max}^{\prime2-\alpha}-\varepsilon_{p,\rm min}^{\prime2-\alpha}}
=({1+z\over \epsilon_\nu})^{2-\alpha}\epsilon_{\rm dec}
\end{equation}
where we assume the spectrum of the ejected protons as $\frac{dN'_p}{d\varepsilon'_p}\propto \varepsilon_{p}^{\prime-\alpha}$,
and $\varepsilon'_{\rm p,min}\sim 2{\rm GeV}$
is the minimum energy of the ejected protons in the rest frame,
and we define a parameter independent on the redshift,
$\epsilon_{\rm dec}\equiv\frac{\varepsilon_{\nu,1}^{2-\alpha}-\varepsilon_{\nu,2}^{2-\alpha}}{\varepsilon_{p,\rm max}^{\prime2-\alpha}-\varepsilon_{p,\rm min}^{\prime2-\alpha}}$.
Adopting an efficiency factor $\eta=0.05-0.15$
for the conversion of ejecta kinetic energy into the relativistic CR protons energy
\cite{Becker2009,Hillas2005},
the total energy of the CR protons is
$E_{\rm CR}=\eta E_{\rm HN}$.
Hereafter, we take the typical kinetic energy of the hypernova
as $E_{\rm HN}=2\times10^{52}{\rm erg}$ and
$\eta=0.1$.
Adopting $\alpha\sim2.1$,
for the neutrinos with energies $0.5-5$ PeV,
we have $\epsilon_{\rm dec}\simeq 0.07$.
The total energy of the observed neutrinos from $\varepsilon_{\nu,1}$ to $\varepsilon_{\nu,2}$ produced by each hyperonva in the ULIRGs is estimated as
\begin{equation}
E_{\nu}\approx 4\times10^{48}{\rm erg}\frac{E_{\rm HN}}{2\times10^{52}{\rm erg}}\frac{\eta}{0.1}\frac{\epsilon_{\rm dec}}{0.07}
\left(\frac{\epsilon_{\nu}}{0.05}\right)^{\alpha-1}\left(\frac{1+z}{3}\right)^{2-\alpha}.
\end{equation}

The produced neutrinos have the similar spectrum as the ejected protons,  i.e., the observed neutrinos spectrum is
$\frac{dN_\nu}{d\varepsilon_\nu}=N_{\rm c}\varepsilon_\nu^{-\alpha}$ \cite{Kelner2006},
then the normalized coefficient of the neutrino spectrum can be calculated
via
\begin{equation}
N_{\rm c}=\frac{E_{\nu}(2-\alpha)}
{\varepsilon_{\nu,2}^{2-\alpha}-\varepsilon_{\nu,1}^{2-\alpha}}
=AE_{\nu},
\end{equation}
where we define a parameter $A\equiv\frac{2-\alpha}{\varepsilon_{\nu,2}^{2-\alpha}-\varepsilon_{\nu,1}^{2-\alpha}}$ to simplify the expression.
Consequently, the diffuse PeV neutrino flux integrating
from local to the high redshift $z$ reads
\begin{eqnarray}
F_{\nu}(\varepsilon_\nu)&=&\frac{1}{4\pi}\frac{c}{H_0}\int_{0}^{z} dz\frac{4\pi D_{
\rm c}(z)^2R_{\rm HN}(z)N_{\rm c}\varepsilon_{\nu}^{2-\alpha}
}
{4\pi D_{\rm L}(z)^2\sqrt{\Omega_{\rm M}(1+z)^3+\Omega_{\Lambda}}}\\\nonumber
&=&
\varepsilon_{\nu}^{2-\alpha}A
\epsilon_{\rm dec}\eta\epsilon_{\nu}^{\alpha-1}E_{\rm HN}\\\nonumber
&&\times\frac{c}{H_0}\int_{0}^{z} dz\frac{R_{\rm HN}(z)}{(1+z)^\alpha\sqrt{\Omega_{\rm M}(1+z)^3+\Omega_{\Lambda}} }
\end{eqnarray}
where the luminosity distance $D_{\rm L}=(1+z)D_{\rm c}$,
while for the specified case
with $\alpha=2$, it reads
\begin{eqnarray}
F_{\nu}(\varepsilon_\nu)
&=&
A^{\ast}\epsilon^{\prime\ast}_{\rm dec}\eta\epsilon_{\nu}E_{\rm HN}
\\\nonumber
&&\times\frac{c}{H_0}\int_{0}^{z} dz\frac{R_{\rm HN}(z)}{(1+z)^2\sqrt{\Omega_{\rm M}(1+z)^3+\Omega_{\Lambda}} }
\end{eqnarray}
with $A^{\ast}=\frac{1}{ln\varepsilon_{\nu,2}-ln\varepsilon_{\nu,1}}$ and
$\epsilon^{\prime\ast}_{\rm dec}=\frac{ln\varepsilon_{p,2}^{\prime}-ln\varepsilon_{p,1}^{\prime}}{ln\varepsilon_{p,\rm max}^{\prime}-ln\varepsilon_{p,\rm min}^{\prime}}$.

In figure 1.,  we show the flux of the diffuse neutrinos, at the energy of
 $1$PeV, from ULIRGs, GRBs and AGNs, for the ejected protons spectrum
$\frac{dN'_p}{d\varepsilon'_p}\propto\varepsilon_p^{\prime-2}$.
The diffuse PeV neutrino flux contributed by ULIRGs from local to $z=2.3$
is $F_{\nu,\rm PeV}\simeq2\times10^{-9} {\rm GeV\,cm^{-2}\,s^{-1}sr^{-1}}$ for $\alpha=2$,
which is comparable with the flux of atmosphere PeV neutrinos,
implying an un-ignorable contribution to the PeV neutrino flux.
However, the detection prospect of PeV neutrinos from ULIRGs is not promising.

We estimate the amount of the detection rate of neutrinos in the energy
range $(\varepsilon_{\nu,1},\varepsilon_{\nu,2})$ via
\begin{eqnarray}
\label{IntNnu}
N_{\nu}&=&
A\epsilon_{\rm dec}\eta\epsilon_{\nu}^{\alpha-1}E_{\rm HN}
\int_{\varepsilon_{\nu,1}}^{\varepsilon_{\nu,2}}
A_{\rm exp}(\varepsilon_\nu)\varepsilon_{\nu}^{-\alpha}d\varepsilon_{\nu}\\\nonumber
&&\times\frac{c}{H_0}\int_{0}^{z} dz\frac{R_{\rm HN}(z)}{(1+z)^\alpha\sqrt{\Omega_{\rm M}(1+z)^3+\Omega_{\Lambda}} },
\end{eqnarray}
where $A_{\rm exp}(\varepsilon_{\nu})$, varying with
the energy of neutrinos, is the exposure coefficient of the detector for the diffuse neutrinos,
with the unit of $\rm cm^2~s~sr$ for $\alpha>2$.
While for $\alpha=2$ we have
\begin{eqnarray}
\label{IntNnu}
N_{\nu}&=&
A^{\ast}\epsilon^{\prime\ast}_{\rm dec}\eta\epsilon_{\nu}E_{\rm HN}
\int_{\varepsilon_{\nu,1}}^{\varepsilon_{\nu,2}}
A_{\rm exp}(\varepsilon_\nu)\varepsilon_{\nu}^{-2}d\varepsilon_{\nu}\\\nonumber
&&\times\frac{c}{H_0}\int_{0}^{z} dz\frac{R_{\rm HN}(z)}{(1+z)^2\sqrt{\Omega_{\rm M}(1+z)^3+\Omega_{\Lambda}} }.
\end{eqnarray}
Considering the most sensitive neutrino detector nowadays, i.e.,
the completed 86 strings IceCube observatory, adopting the effective
area varying with the energy of protons,
equation (\ref{IntNnu})
gives $N_{\nu}=0.1$
for one year observation.

\begin{figure}
\includegraphics[width=95mm,angle=0]{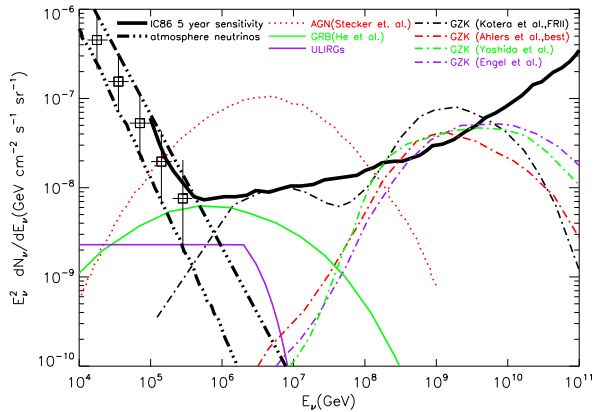}
\caption{The flux of the diffuse neutrino emission
from ULIRGs (purple solid line), GRBs (green solid line \cite{He2012}),
AGNs (red dotted line \cite{Stecker2007,Stecker2007APh}),
assuming that the spectrum of the ejected protons is
${dN'_p}/{d\varepsilon'_p}\propto\varepsilon_p^{\prime-2}$.
The black, red, green and purple dash-dotted lines represent the
Greisen-Zatsepin-Kuzmin (GZK) neutrinos \cite{Greisen1966, Zatsepin1966} referring to the models in \cite{Kotera2010} (among Faranoff-Riley type II galaxies, i.e., FRII),
\cite{Ahlers2010} (with the best parameters that fit the cosmic ray data), \cite{Yoshida1993}
and \cite{Engel2001},
respectively.
The black thick solid line represents
the sensitivity of IceCube 86 strings for 5 years. The atmospheric neutrinos are presented by the
data with error bars, which is measured by IceCube \cite{IceCube2011PRD}.
The two black dash-triple-dotted lines  are the upper bound and lower bound of the atmosphere neutrinos
extrapolating to the high energy. \\
}
\label{fig:Cartoon}
\end{figure}

\section{Conclusion and Discussion.}
ULIRGs are a group of galaxies with ultra-luminous Infrared emission
($L_{\rm IR}>10^{12}L_{\odot}$) and
a high SFR ($\gtrsim 200 M_{\odot} {\rm yr^{-1}}$).
Consequently, a high hypernovae rate,
$R_{\rm HN}\sim1000{\rm yr^{-1}Gpc^{-3}}$ at redshift $1.2\lesssim z\lesssim2.3$
is expected.
Since the hypernovae can drive energetic shocks and are able
to accelerate protons to $10^{17}$ eV,
we propose that huge amounts of protons with spectrum
$\frac{dN'_p}{d\varepsilon'_p}\propto\varepsilon_p^{\prime-\alpha}$ are
ejected into these ULIRGs.
The observations indicate that ULIRGs have a very
high gas surface density, therefore
the protons is expected to lose most of their energy through interacting
with the dense ISM in ULIRGs before escaping,
providing a un-negligible contribution to the PeV neutrinos flux
($\sim 2\times10^{-9}{\rm GeV\,cm^{-2}\,s^{-1}sr^{-1}}$). Its flux comparing with that of the atmosphere neutrinos, the GRB neutrinos and the AGN neutrinos have been presented in Fig.1.
The ULIRG neutrino component is likely characterized by a cutoff (or break) at $\sim$ a few PeV
since the hypernovae are likely only able to accelerate the CR protons up to
$\sim100$ PeV and the ULIRGs can not confine the protons with
energy much larger than $\sim100$ PeV, either. Such a component may be detected in 20 years by the IceCube full configuration.

Since the diffuse neutrino emission from ULIRGs are expected to be much rarer (see Fig.1),
finally we suggest that the two PeV neutrino candidates reported by IceCube Collaboration \cite{Ishihara2012},
if cosmological, may be from other energetic sources, such as AGNs, GRBs (see \cite{Cholis2012}, but see \cite{He2012,Liu2012}), and cosmogenic neutrinos (see \cite{Barger2013}, but see \cite{Bhattacharya2012,Roulet2013}). The origins of the reported two PeV neutrinos are
highly controversial so far,
we anticipate more observations from the IceCube to draw a firm conclusion in the future.


{\it Acknowledgments.} HNH thanks Ruoyu Liu for the useful discussion and Shigehiro Nagataki for the useful suggestion. This work was supported in part by 973 Program of China under grants 2009CB824800 and 2013CB837000, National Natural Science of China under grants 11173064 and 11273063, and by China Postdoctoral science foundation under grant 2012M521137.  YZF is also supported by the 100
Talents program of Chinese Academy of Sciences and the Foundation for
Distinguished Young Scholars of Jiangsu Province, China (No. BK2012047).
SML is also supported by the Recruitment Program of Global Experts from the Central Organization Committee.

$^\ast$Corresponding author.\\
Electric addresses: hnhe@pmo.ac.cn, taowang@nju.edu.cn, yzfan@pmo.ac.cn, liusm@pmo.ac.cn, dmwei@pmo.ac.cn

\end{document}